\documentclass[12pt,twoside,fleqn]{article}
\usepackage{espcrc1}

\newcommand{\AmS}{{\protect\the\textfont2
  A\kern-.1667em\lower.5ex\hbox{M}\kern-.125emS}}

\title{\hspace{12cm}{\bf NT@UW-99-49}\\Light quark distributions in the proton sea}
\author{Mary Alberg\address{Department of Physics, Seattle University,  
       Seattle, WA 98122, USA and \\
	Department of Physics, University of Washington, Seattle, WA 98195, USA }%
 \thanks{Contributions to the work described in this talk have been made by
students Thomas Falter (University of Giessen) and Adam Graunke (Seattle University).}       
        and 
       Ernest M. Henley\address{Department of Physics, University of Washington,
Seattle, WA 98195, USA and
 \\Institute for Nuclear Theory, University of Washington, Seattle, WA 98195, USA }}
       
\begin{document}

\maketitle

\begin{abstract}

We use the meson cloud model
 to calculate  $\bar{d}(x) - \bar{u}(x)$ and $
\bar{d}(x)/\bar{u}(x)$ in the proton. We show that a modification of the
symmetric,  perturbative part of the light quark sea provides better agreement with
the ratio $
\bar{d}(x)/\bar{u}(x)$.
\end{abstract}

\section{Introduction}

Flavor asymmetry in the light quark sea of the proton has been well-established by
experiment \cite{NMC,NA51,E866,HERMES}. The violation of the Gottfried sum
rule found by NMC \cite{NMC} indicated that 
$D\equiv   
     \int_0^1 dx\; (\bar{d}(x)-\bar{u}(x))=0.148\pm0.039$.
In Drell-Yan experiments, $\bar{d}/\bar{u}$ was determined to be
greater than 2 at
$x=.18$ by NA51 \cite{NA51}, and the $x$-dependence of this ratio has recently been
measured by E866 \cite{E866} in the region $0.02 \leq x \leq 0.345$. From their data and
global parton distributions \cite{PDF} E866 also determined the difference $\bar{d}(x) -
\bar{u}(x)$ and $D=0.100\pm0.018$. Flavor asymmetry consistent with
these results has also been seen in the measurement of
$\bar{d}(x) -
\bar{u}(x)$ by HERMES
\cite{HERMES}. 

It was first suggested by
Thomas \cite{AWT83}, and later by Henley and Miller \cite{HM90} that a natural
explanation for this asymmetry is the meson cloud of the proton. The net positive
charge of the cloud leads to an excess of $\bar{d}$ over $\bar{u}$. Other causes
for the asymmetry have been invoked, such as antisymmetrization \cite{FF77,MST98},
but these have been insufficient to describe the data. For  reviews, see
Refs.~\cite{ST98,sK98}. 

We wish to emphasize that the E866 measurements provide a
critical test of our understanding of the flavor-symmetric (FS) contributions to
the light quark sea, as well as the flavor-asymmetric (FA) contributions. Meson
cloud models \cite{MST98,sK91,SST91,KFS96,NSSS99} provide a reasonably good
description of the
$\bar{d}(x) - \bar{u}(x)$  asymmetry, which depends on FA contributions alone, but fail to
explain the broad maximum in the ratio
$\bar{d}(x)/\bar{u}(x)$ at $x \approx 0.18$ and its return to unity, or even lower
values, at larger $x$. FS terms also contribute to this ratio 
\begin{equation}
\frac{\bar{d}(x)}{\bar{u}(x)} = \frac{\bar{d}(x)-\bar{u}(x)}{\bar{u}(x)}+1,
\end{equation}
and it is clear that $\bar{u}(x)$ falls too rapidly with $x$ in these models. We
have proposed \cite{AFH98} that agreement with data can be improved by using harder
distributions for the perturbative contributions to $\bar{u}(x)$, motivated by
their origin in gluon splitting.    
\section{Pion cloud model}
To illustrate this argument we use a pion cloud model. Other states should be
included in a full calculation, but e.g. the  
$\rho$ meson increases $\bar{d}(x) - \bar{u}(x)$, whereas the intermediate 
$\Delta$ decreases it, so these effects tend to cancel. 
The wave function of the proton is
written in terms of a Fock state expansion
\begin{eqnarray}
\mid p\rangle = \sqrt{Z}\mid p\rangle_{\rm bare} + \sum_{MB}\int dy\;
d^2\vec{k}_\perp\; \phi_{BM}(y,k_\perp^2)
\mid  B(y,\vec{k}_\perp) M(1-y, - \vec{k}_\perp)\rangle \;,
\end {eqnarray}
with $BM = p\pi^0, n\pi^+$. The factor $\sqrt{Z}$ is a wavefunction renormalization
constant and 
$\phi_{BM}(y,k_\perp^2)$ is the probability amplitude for finding a physical
nucleon in a state consisting of a baryon $B$ with longitudinal momentum fraction
$y$, transverse momentum $\vec{k}_\perp$, and a meson 
$M$ of 
momentum fraction $(1-y)$, transverse momentum $-\vec{k}_\perp$.
The quark distribution functions $q(x)$ in the proton are given by
\begin{equation}
q(x) = q^{\rm bare}(x) + \delta q(x) \; ,
\end{equation}
with 
\begin {eqnarray}
\delta q(x) = \sum_{MB}
\left(\int_x^1 f_{MB}(y) q_M (\frac{x}{y})\frac{dy}{y} \; 
+\int_x^1 f_{BM}(y) q_B(\frac{x}{y})\frac{dy}{y}\right),
\end{eqnarray}
\begin{equation}
f_{BM}(y) = \int_0^\infty\mid\phi_{BM}(y,k_\perp^2)\mid^2~d^2k_\perp\;,
\end{equation}
and
\begin{equation}
f_{MB}(y) = f_{BM}(1-y) \; .\label{sym}
\end{equation}
The splitting function $f_{n\pi^+}(y)=2f_{p\pi^0}(y)$, with \cite{ST98,Holt}
\begin{equation}
f_{p\pi^0}(y)=\frac{g^2}{16\pi^2}\frac{1}{y^2(1-y)}\int_0^\infty
dk_\perp^2\vert
G_{\pi}(y,k_\perp^2)\vert^2\frac{m_N^2(1-y)^2
+k_\perp^2}{[m_{N}^2-M_{N\pi}^2(y,k_\perp^2)]^2},
\end{equation}
in which $M_{N\pi}^2(y,k_\perp^2)$ is the invariant mass squared of the
intermediate Fock state
\begin{equation}
M_{N\pi}^2(y,k_\perp^2)=\frac{m_N^2+k_\perp^2}{y}+\frac{m_\pi^2+k_\perp^2}{1-y}.
\end{equation}
We use an exponential form for the cutoff
\begin{equation}
G_{\pi}(y,k_\perp^2)=\exp(\frac{m_{N}^2-M_{N\pi}^2(y,k_\perp^2)}{2\Lambda^2})
\end{equation}
which insures that the identity (\ref{sym}) is satisfied \cite{ST98,SEHS96}.
We use Holtmann's parametrization \cite{Holt} of the bare nucleon symmetric sea
($\bar{Q}_{\rm bare}=u_{\rm sea}=\bar{u}_{\rm sea}=d_{\rm sea}=\bar{d}_{\rm sea}$)
\begin{eqnarray}
x\bar{Q}_{\rm bare}(x)=0.11(1-x)^{15.8} \label{soft}  .
\end{eqnarray}
\begin{figure}[t]
\vspace{4.4cm}
\includegraphics{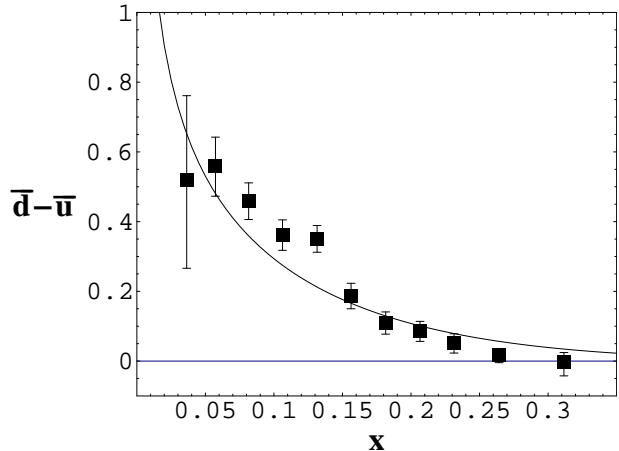}
\caption{Comparison of our pion cloud model with data \cite{E866} for 
$\bar{d}(x) - \bar{u}(x)$. The 
cutoff constant  $\Lambda = 0.83 $ GeV, for which 
$D=\int_0^1 (\bar{d}(x) - \bar{u}(x)) dx =0.100$.}
\label{fig:dbar-ubar}
\end{figure}
Since gluon splitting is the origin of this sea, we also use a harder
distribution for the bare sea quarks, of the form found in a recent
determination of the gluon distribution \cite{Vogt}
\begin{equation}
x\bar{Q}^{'}_{\rm bare}(x)=0.0124 x^{-0.36}(1-x)^{3.8}. \label{hard}
\end{equation} 
For the pion valence quarks $q_v$ and sea quarks $q_{\rm sea}$ 
 we use \cite{SMRS92}
\begin{eqnarray}
x q_v(x)=0.99x^{0.61}(1-x)^{1.02}, &    &  
xq_{\rm sea}(x)=0.2(1-x)^{5.0}.
\end{eqnarray}
The $\pi$-nucleon coupling constant  is taken as ${g_\pi^2\over 4\pi}=13.6$
The value of $\Lambda = 0.83 $ GeV is chosen
  to reproduce the integrated asymmetry $D=0.100$  \cite{E866}.
\section{Discussion} 
The results of our calculations for $\bar{d}(x)-\bar{u}(x)$
are shown in Fig.~1, and
those for the ratio $\bar{d}(x)/\bar u(x)$ are shown in Fig.~2.
In Fig. 1 the flavor asymmetry is caused entirely by the $\pi^+$, and our result is not
affected by the different forms we have chosen for the bare sea FS contribution.  In Fig.~2
the solid curve shown is for the perturbative sea quark distribution of (\ref{soft}). The
dashed curve is for the gluonic form of (\ref{hard}). 
\begin{figure}[t]
\vspace{3.8cm}
\begin{center}
\includegraphics{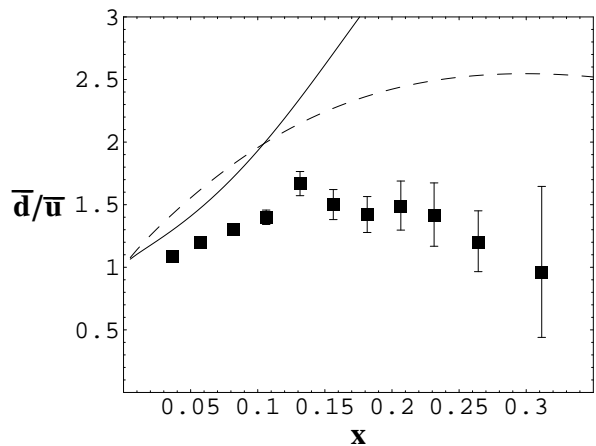}
\end{center}
\caption{Comparison of our pion cloud model with data \cite{E866} for $\bar{d}(x)/
\bar{u}(x)$. The  solid curve is for the perturbative sea quark distribution of
(\ref{soft}). The dashed curve is for the gluonic form of (\ref{hard}).}
\label{fig:dbar/ubar}
\end{figure}
It is clear from Fig. 2  
that a better description of the present data is provided by using a harder
distribution for the symmetric sea. Of course other FS contributions
could produce a similar improvement in agreement between theory and experiment. We
have recently examined the role of the $\omega$ in the meson cloud of the proton
and find this to be the case \cite{AHM99}. The
$\sigma$ meson would also tend to suppress  $\bar{d}/ \bar{u}$.
A complete calculation must include all the components of the meson cloud.

Forthcoming analyses of new E866 data \cite{theses} and proposed experiments
\cite{prop} will further test our models of both perturbative and non-perturbative
contributions to the flavor-symmetric and flavor-asymmetric components of the proton sea.

This work has been supported in part by the U.S. Department of Energy and the
Research in Undergraduate Institutions program of the U.S. National Science
Foundation (Grant No. 9722023). We thank the
members of the E866 collaboration for helpful discussions of their data and analyses.
One of us (EMH) thanks J. Speth for a useful discussion and for his hospitality at
the KFZ in Juelich.

\end{document}